\documentclass[aps,prx,twocolumn,superscriptaddress,showpacs]{revtex4}

\usepackage{eurosym}
\usepackage{amsfonts}
\usepackage{amssymb}
\usepackage{amsmath}
\usepackage{graphicx}
\usepackage{color}
\begin{document}
\title{
Enhanced superconductivity by strain and carrier-doping in borophene: A first principles prediction
}

\author{R. C. Xiao}
\email[The authors contributed equally to this work.]{}
\affiliation{Key Laboratory of Materials Physics, Institute of Solid
State Physics, Chinese Academy of Sciences, Hefei 230031, China}
\affiliation{University of Science and Technology of China, Hefei, 230026, China}

\author{D. F. Shao}
\email[The authors contributed equally to this work.]{}
\affiliation{Key Laboratory of Materials Physics, Institute of Solid
State Physics, Chinese Academy of Sciences, Hefei 230031, China}

\author{W. J. Lu}
\email{wjlu@issp.ac.cn} 
\affiliation{Key Laboratory of Materials
Physics, Institute of Solid State Physics, Chinese Academy of
Sciences, Hefei 230031, China}

\author{H. Y. Lv}
\affiliation{Key Laboratory of Materials Physics, Institute of Solid
State Physics, Chinese Academy of Sciences, Hefei 230031, China}

\author{J. Y. Li}
\affiliation{Key Laboratory of Materials Physics, Institute of Solid
State Physics, Chinese Academy of Sciences, Hefei 230031, China}
\affiliation{University of Science and Technology of China, Hefei, 230026, China}

\author{Y. P. Sun}
\email{ypsun@issp.ac.cn}
\affiliation{High Magnetic Field Laboratory, Chinese Academy of
Sciences, Hefei 230031, China}
\affiliation{Key Laboratory of Materials Physics, Institute of Solid
State Physics, Chinese Academy of Sciences, Hefei 230031, China}
\affiliation{Collaborative Innovation Center of Microstructures,
Nanjing University, Nanjing 210093, China }

%===============================abstract============================%
\begin{abstract}
We predict by first principles calculations that the recently prepared borophene is a pristine two-dimensional (2D) monolayer superconductor, in which the superconductivity can be significantly enhanced by strain and charge carrier doping. The intrinsic metallic ground state with high density of states at Fermi energy and strong Fermi surface nesting lead to sizeable electron-phonon coupling, making the freestanding borophene superconduct with $T_c$ close to 19.0 K. The tensile strain can increase $T_c$ to 27.4 K, while the hole doping can notably increase $T_c$ to 34.8 K. The results indicate that the borophene grown on substrates with large lattice parameters or under photoexcitation can show enhanced superconductivity with $T_c$ far more above liquid hydrogen temperature of 20.3 K, which will largely broaden the applications of such novel material.
\end{abstract}
\pacs{74.78.-w, 71.18.+y, 73.22.-f}
\maketitle
%===================================================================%
\section{Introduction}
Since the discovery of graphene \cite{science_666-669,nature_197-200}, two-dimensional (2D) monolayer materials have gained significant attentions. Monolayer transition metal dichalcogenide MoS$_2$ \cite{Nature_nanotechnology_699-712,Andrea-nanolett-2010,PRL-Mak-2010}, \textit{h}-BN \cite{Science_217-220,Pacile-APL-2008,Song-nanolett-2010}, silicene \cite{Applied_Physics_Letters_183102,Applied-Physics-Letters-223109,Vogt-2012-PRL}, phosphorene \cite{Nano_Research_853-859,Nat_Nanotechnol_372-7}, etc. were successfully prepared. Novel properties emerge in those monolayer materials, implying the promising applications in the future electronic devices. However, among the reported remarkable properties of the most of the prepared pristine monolayer materials, superconductivity was rarely found. It is due to the intrinsic semi-metal or semiconductor ground states of such materials. The vanishing density of states (DOS) at Fermi energy ($E_F$) prevents the emergence of superconductivity. A monolayer material showing superconductivity can be very useful in the nanoscale superconductor devices \cite{Nat_Nano_703-711,Science_1045-1048,Applied-Physics_Letters_222506,Phys_Rev_Lett_027203,Physical_Review_B_134530} to achieve single-spin sensitivity for measuring and controlling. Lots of predictions suggested that such superconductivity can be introduced by the metallization of those monolayer materials such as Li atom adsorption on graphene \cite{Nature_Physics_131-134,Guzman-2D-2014}, hole doping in fully hydrogenated graphene \cite{Physical_review_letters_037002}, and electron doping in graphene under strain \cite{Phys-Rev-Lett-196802}, monolayer MoS$_2$ \cite{Physical_Review_B_241408}, siliene \cite{EPL_(Europhysics_Letters)_36001} and phosphorene \cite{EPL_(Europhysics_Letters)_67004,New_Journal_of_Physics_035008}. However, in experiment it seems that only the superconductivity in Li-adsorbed graphene was verified \cite{Ludbrook-PNAS-2015}. The reason why other predictions are still not verified might be the difficulty of sample preparation. The pristine monolayer materials with intrinsic metallic states should be more promising to realize such superconductivity.

Very recently, borophene, the 2D monolayer boron, was successfully prepared by different groups\cite{Science_1513-1516,arXiv_1512.05029,Science_1468-1469}. Unlike the most reported monolayer materials mentioned above, borophene was not prepared by exfoliating from the bulk layered materials, but was directly grown on some metal substrates. Different structures of borophene were reported by different groups, implying that the structures of borophene strongly depend on the preparation conditions. Remarkable properties such as anisotropic conductivity, negative Poisson¡¯s ratio, high Young's modulus in \emph{\textbf{a}} direction potentially exceeding the value of graphene \cite{Science_1513-1516}, and potential anisotropic Dirac fermion in hydrogenated borophene \cite{arXiv_1602.03620} were suggested in borophene, indicating the promising applications of the novel monolayer material. Considering the intrinsic metallic behavior with high DOS at $E_F$ ($N(E_F)$) verified by different reports \cite{Science_1513-1516,arxiv-1601,Penev-nano-lett-2016}, one can expect for the possibility of superconductivity in this pristine monolayer material. Recently Gao \textit{et al.} \cite{arXiv_1602.02930} and Penev \textit{et al.} \cite{Penev-nano-lett-2016} respectively predicted the phonon-mediated superconductivity with $T_c$ of 10$\sim$20 K for the reported structures with vacancy fraction \textit{x} = 0, 1/5, 1/6 and 1/3 \cite{Science_1513-1516,arxive-1512.0527,Arxiv-01393,Angewandte_Chemie_13214-13218,Penev-nano-lett-2016}.

Here we further demonstrate by first principles calculations that the superconductivity in borophene with vacancy-free structure (\textit{x} = 0) \cite{Science_1513-1516}, which was estimated to have the highest $T_c$ compared with those with vacancy structures \cite{Penev-nano-lett-2016,arXiv_1602.02930}, could be significantly enhanced by strain and carrier doing. The tensile strain can increase $T_c$ to 27.4 K, while the hole doping can increase $T_c$ to 34.8 K, which are far more above the liquid hydrogen temperature of 20.3 K. Our predictions suggest that the superconductivity in borophene grown on a substrate with large lattice parameters, or under photoexcitation could be significantly enhanced, which will largely broaden the applications of such novel material.

\begin{figure}[h]
\begin{flushleft}
\includegraphics[width=0.99\columnwidth]{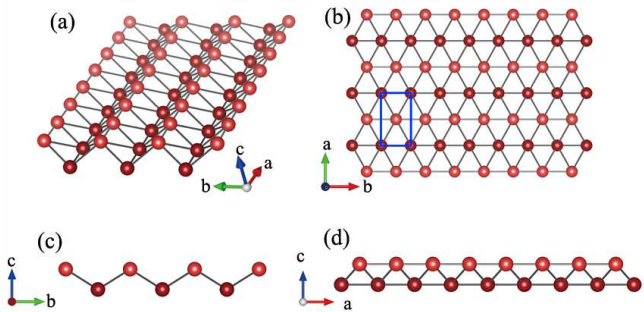}
\caption{Crystal structure of borophene. (a)-(d) show the structure view from different directions. The unit cell is denoted with blue rectangle in (b).}
\end{flushleft}
\end{figure}

\section{Methods}
The calculations based on density functional theory (DFT) were carried out using QUANTUM-ESPRESSO package \cite{Journal_of_Physics_Condensed_Matter_395502}. The ultrasoft pseudo-potentials \cite{Physical_Review_B_7892} and the generalized gradient approximation (GGA) according to the Perdew-Wang 91 gradient-corrected functional \cite{Phys.-Rev.-B-6671-6687} were used. To simulate the monolayer, a vacuum layer of $15\ \mathrm{\AA}$ was used. The energy cutoff for the plane wave basis was set to 35 Ry. The Brillouin zone was sampled with a $32\times20\times1$ mesh of k-points. The Vanderbilt-Marzari Fermi smearing method with a smearing parameter of $\sigma=0.02$ Ry was used for the calculations of the total energy and the electron charge density. The phonon dispersion and electron-phonon coupling constants were calculated using density functional perturbation theory (DFPT) \cite{Reviews_of_Modern_Physics_515-562} with a $16\times10\times1$ mesh of q-points. The double Fermi-surface averages of electron-phonon matrix elements were calculated using $96\times96\times1$ k-points.

\begin{figure}[h]
\includegraphics[width=0.8\columnwidth]{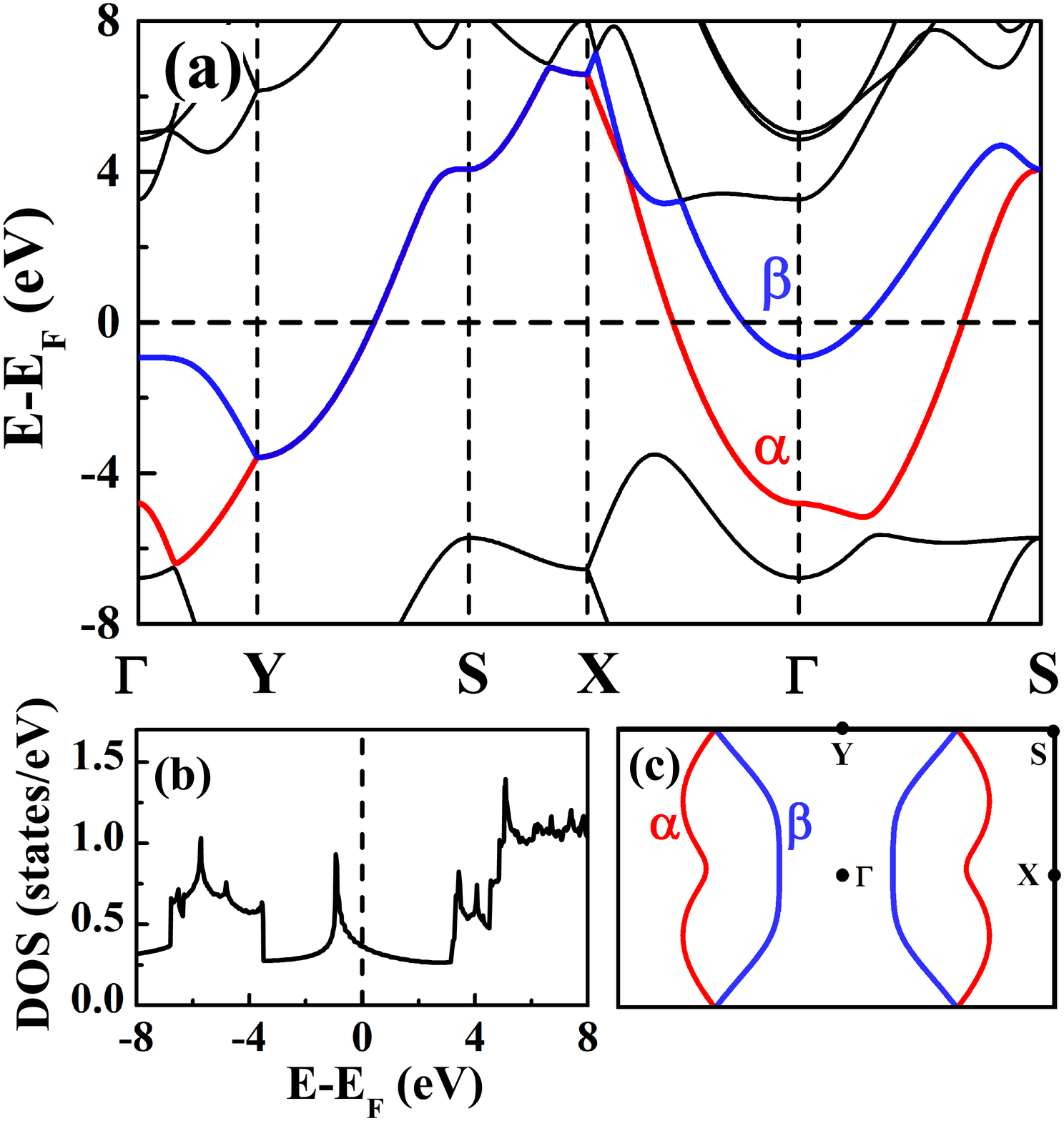}
\caption{Electronic properties of borophene: (a) band structure, (b) density of states, (c) Fermi surfaces. The bands crossing $E_F$ and their related Fermi surfaces are denoted as $\alpha$ (red color) and $\beta$ (blue color).}
\end{figure}

\section{Results and discussions}
Our investigation starts from a hypothetic freestanding borophene. As shown in Fig. 1, borophene has a highly anisotropic crystal structure with a rectangular unit cell. There are corrugations along \emph{\textbf{b}} direction, while no pucker in \emph{\textbf{a}} direction is found. The fullly optimized lattice parameters of freestanding borophene are $\emph{a}=1.611 \mathrm{\AA}$ and $\emph{b}=2.872\mathrm{\AA}$, which are smaller than the reported experimental values \cite{Science_1513-1516}, indicating the real lattice parameters are strongly dependent on the substrates.

Figure 2(a) shows the calculated band structure, which is agreement well with the previous calculations \cite{Science_1513-1516,arxiv-1601,Penev-nano-lett-2016}. We denote the two bands crossing $E_F$ as $\alpha$ and $\beta$ bands respectively. One can note that both these two bands do not cross $E_F$ in the $\Gamma-Y$ and $S-X$ directions, which are parallel to the \emph{\textbf{b}} direction, indicating the strong highly anisotropic electronic properties. Clearly, the out-of-plane corrugations along the \emph{\textbf{b}} direction open a band gap along the $\Gamma-Y$ and $S-X$ directions. The two bands form the Fermi sheets symmetrical to $\Gamma-Y$ (Fig. 2(c)). The calculated DOS is shown in Fig. 2(b). The large $N(E_F)$ of 0.36 states/eV per unit cell indicates a good metallic behavior. A DOS peak locates at about 1 eV below $E_F$, indicating that hole doping should significantly increase $N(E_F)$.
\begin{figure}[h]
\includegraphics[width=0.99\columnwidth]{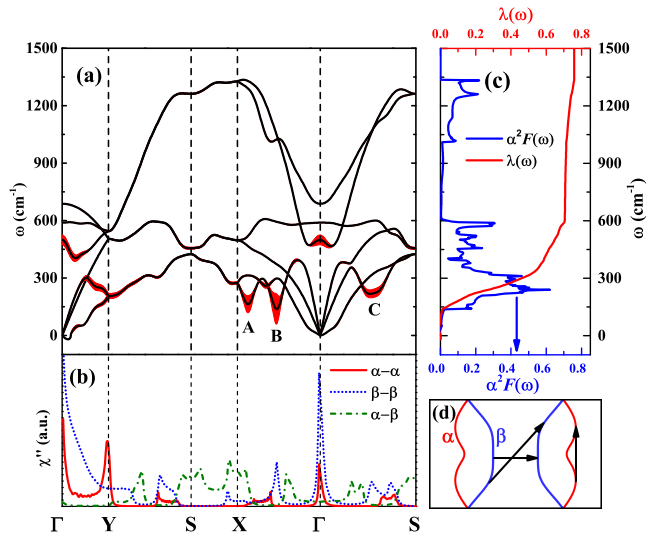}
\caption{(a) Phonon dispersion, (b) imaginary part of electron susceptibility $\chi''$, (c) Eliashberg function $\alpha^2F(\omega)$ and integrated electron-phonon coupling strength $\lambda$, and (d) possible nesting vectors of borophene. The phonon dispersion in (a) is decorated with red ribbons, proportional to the partial electron-phonon coupling strength $\lambda_{\mathbf{q} v}$.}
\end{figure}

Figure 3(a) shows the calculated phonon dispersion of borophene based on DFPT. We found the instability is very close to $\Gamma$, which was also reported by the previous calculations \cite{Science_1513-1516,arxiv-1601,arXiv_1602.00456}. Such instability is consistent with the instability against long-wavelength transversal waves, which is suggested to be fixed by defects, such as ripples or grain boundaries \cite{Science_1513-1516}. Besides, we found three notable softened acoustic modes (denoted with A, B, and C in Fig. 3(a)). The softened mode A close to X point was observed in the previous calculations \cite{Science_1513-1516,arxiv-1601,Penev-nano-lett-2016,arXiv_1602.00456}. However, the mode B located at the approximate center of $\Gamma-X$ was absent in the previous calculations using supercell method \cite{Science_1513-1516,arxiv-1601,arXiv_1602.00456}. Moreover, very recent calculations by DFPT show the imaginary frequencies of modes A and B \cite{Penev-nano-lett-2016}. Such differences might be due to the different methods and pseudo-potentials used in the phonon calculations. The stability of the phonon modes we calculated permits us to estimate the electron-phonon coupling and the superconductivity in borophene accurately.

We decorate the calculated electron-phonon coupling strength at each mode ($\lambda_{\mathbf{q} v}$) in phonon dispersion in Fig. 3(a), which is defined by
\begin{equation}
\lambda_{\mathbf{q} v}=\frac{\gamma_{\mathbf{q} v}}{\pi \hbar N(E_F) \omega_{\mathbf{q} v}^2} .
%==================equation1
\end{equation}
The phonon line width $\gamma_{\mathbf{q} v}$ is defined by
\begin{equation}
\begin{split}
\gamma_{\mathbf{q} v}=2\pi \omega_{\mathbf{q} v}\sum_{ij}\int\frac{\mathrm{d}\mathbf{k}}{\Omega_{BZ}}
|g_{\mathbf{q}v}(\mathbf{k},i,j)|^2 \\ \times \delta(\epsilon_{\mathbf{k},i}-E_F)\delta(\epsilon_{\mathbf{k+q},j}-E_F) ,
%==================equation2
\end{split}
\end{equation}
where $g_{\mathbf{q}v}(\mathbf{k},i,j)$ is the electron-phonon coupling matrix element. Since Fermi surface nesting can be reflected by the imaginary part of electron susceptibility
\begin{equation}
\chi''(\mathbf{q})=\sum_{\mathbf{q}}\delta({\epsilon_\mathbf{k}-E_F})\delta({\epsilon_{\mathbf{k+q}}-E_F}) .
%==================equation3
\end{equation}
According to such definitions (Eqs. (1)-(3)), the strong Fermi nesting will enhance the electron-phonon coupling significantly. As shown in Fig. 3(a), the softened modes A, B and C predominately contribute to the electron-phonon coupling. At the locations of the modes B and C there are sharp $\chi''$ peaks of $\beta$-$\beta$ band (Fig. 3(b)), indicating that the strong electron-phonon coupling is contributed by the nesting between the Fermi sheets formed by band $\beta$-$\beta$, as shown in Fig. 3(d). On the other hand, at the location of the mode A, no substantial peak of $\chi''$ can be found. Therefore, the electron-phonon coupling contributed by the mode A is mainly from the intrinsic large $g_{\mathbf{q}v}(\mathbf{k},i,j)$. Besides, two acoustic modes around Y point also contribute to the sizeable electron-phonon coupling, which is due to the nesting formed by band $\alpha$-$\alpha$ (Figs. 3(b) and (d)). The contribution from the lowest optic mode around $\Gamma$ is related to the intra-sheet nesting.

The total electron-phonon coupling constant $\lambda$ can be obtained by
\begin{equation}
\lambda=\sum_{\mathbf{q} v}\lambda_{\mathbf{q} v}
=2\int\frac{\alpha^2F(\omega)}{\omega}\mathrm{d}\omega ,
%==================equation4
\end{equation}
where the Eliashberg spectral function is
\begin{equation}
  \alpha^2F(\omega)=\frac{1}{2\pi N(E_F)}\sum_{\mathbf{q}v}
  \delta(\omega-\omega_{\mathbf{q}v})\frac{\gamma_{\mathbf{q}v}}{\hbar \omega_{\mathbf{q}v}}.
%==================equation5
\end{equation}

We plotted $\alpha^2F(\omega)$ and integrated $\lambda$ in Fig. 3(c). One can note that the large contribution of the electron-phonon coupling from modes A, B, C, and acoustic modes around Y point lead to a broaden peak from 100 cm$^{-1}$ to 300 cm$^{-1}$. Such modes contribute about 70\% of the total $\lambda$. The lowest optic mode around $\Gamma$ leads to small peaks of $\alpha^2F(\omega)$.
We estimated the superconducting transition temperature $T_c$ based on the Allen-Dynes formula \cite{Physical_Review_B_905}
\begin{equation}
T_c=\frac{\omega_{log}}{1.2}\exp\left(-\frac{1.04(1+\lambda)}{\lambda-\mu^*-0.62\lambda\mu^*}\right) ,
%====================equation6
\end{equation}
where the Coulomb pseudopotential $\mu^*$ is set to a typical value of 0.1. The logarithmically averaged characteristic phonon frequency $\omega_{log}$ is defined as
\begin{equation}
\omega_{log}=\exp\left(\frac{2}{\lambda}\int\frac{d\omega}{\omega}\alpha^{2}F(\omega)\log\omega\right).
%====================equation7
\end{equation}
The calculated $\lambda$=0.79 and $\omega_{log}$=421.3 K lead to $T_c$=19.0 K. Such estimation indicates that borophene is a pristine monolayer superconductor, which coincides with the previous prediction \cite{Penev-nano-lett-2016}.

\begin{figure*}
\includegraphics[width=0.7\textwidth]{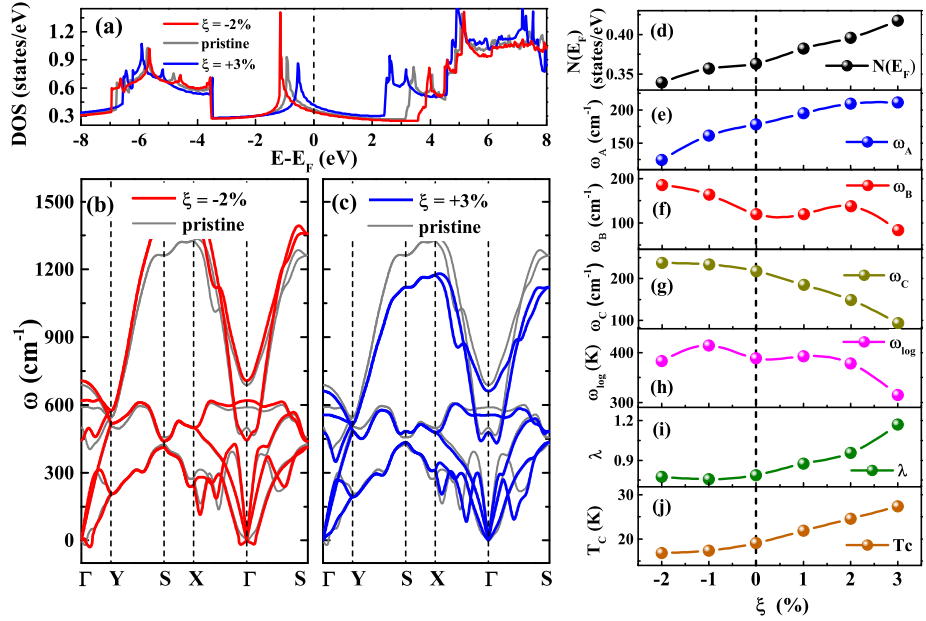}
\caption{(a) DOS of borophene under strain of -2\% (red) and +3\% (blue). DOS of pristine freestanding borophene (grey) is presented for comparison. (b) and (c) are phonon dispersions of such samples, respectively. (d)-(j) show the strain effects on $N(E_F)$, $\omega_A$, $\omega_B$, $\omega_C$, $\omega_{log}$, $\lambda$ and $T_c$ of borophene, respectively.}
\end{figure*}

\begin{figure*}
\includegraphics[width=0.7\textwidth]{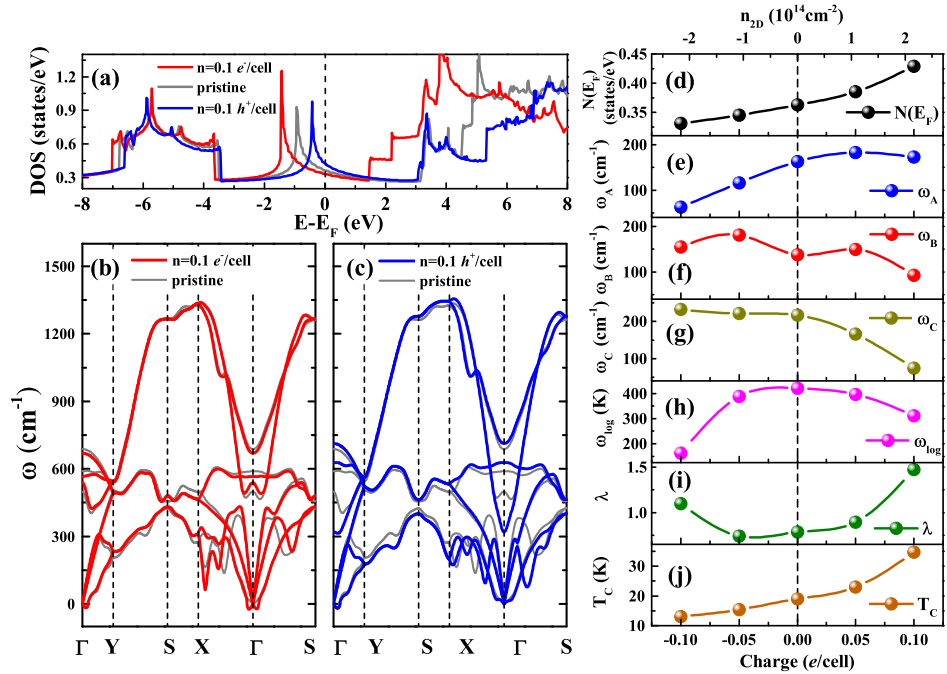}
\caption{(a) DOS of borophene doped by 0.1 electron (red) and 0.1 hole (blue). DOS of pristine freestanding borophene (grey) is presented for comparison. (b) and (c) are phonon dispersions of borophene of such samples, respectively. (d)-(j) show the carrier doping effects on $N(E_F)$, $\omega_A$, $\omega_B$, $\omega_C$, $\omega_{log}$, $\lambda$ and $T_c$ of borophene, respectively.}
\end{figure*}

Since borophene is directly grown on some metal substrates instead of being exfoliated from bulk layered materials, the estimation of freestanding sample is not enough to reflect the reality. Therefore, we calculated the borophene under different strain to simulate the samples grown on the substrates with different lattice parameters. We found that borophene is stable under the strain range of $-2\%\leqslant\xi\leqslant3\%$ (positive value means tensile train, while negative value means compressive strain), which is slightly different with the recent predictions \cite{arXiv_1602.00456}. That might be due to the different calculation methods. Figure 4(a) shows the DOS of borophene under strain of -2\% and +3\%. One can note that the tensile strain moves the DOS peak up toward $E_F$, while the compressive strain moves the DOS peak away from $E_F$. Figures 4(b) and (c) show the phonon dispersions of borophene under strain of -2\% and +3\% respectively. It can be found that the mode A is softened under compressive strain, while the modes B and C are hardened. On the other hand, the tensile strain hardens the mode A and significantly softens the modes B and C. We presented the variations of some superconductivity-related parameters in Figs. 4(d)-(j). Generally speaking, the tensile strain enhances the electron-phonon coupling remarkably and the compressive strain weakens electron-phonon coupling largely. The variation trend of $\lambda$ coincides with that of $N(E_F)$, which is corresponding to BCS theory \cite{Cooper-1189,Physical_Review_162,Physical_Review_1175}. Moreover, according to Eq. (1), a mode with lower frequency will strongly contribute to the electron-phonon coupling. In the present case, the variation of electron-phonon coupling with strain is predominately contributed by the modes B and C, since the changes of the frequencies of such modes are more notable than that of mode A (Figs. 4(e)-(g)). Figure 4(j) shows the estimations of $T_c$ of borophene under different strain. According to the calculations, we can conclude that the substrates with larger lattice parameters will favor the superconductivity. For the sample under tensile strain of 3\%, for which the lattice parameters are very close to the reported experimental values \cite{Science_1513-1516}, $T_c$ of 27.4 K is estimated, which is remarkably higher than the liquid hydrogen temperature of 20.3 K.

In the 2D monolayer systems, charge carrier doping is an easy way to control the electronic properties. Controls by using gate-related methods to introduce electrons and using photoexcitation to dope holes are successfully realized in many 2D systems \cite{Science_1193-1196,Nat_Commun_8826,Science_advances_e1500168,arXiv_1512.06553,Science_177-180}. In the present case, since $N(E_F)$ is closely related to the electron-phonon coupling and can be easily tuned by the carrier-doping which induces the change of $E_F$, we also investigated the carrier doping effects on borophene (Fig. 5). As expected, $N(E_F)$ decreases with the electron doping and increases with the hole doping (Figs. 5(a) and (d)), since the carrier doping changes the distance between $E_F$ and the DOS peak as mentioned above. However, unlike the strain effect, we found that the variation of $\lambda$ is not same as that of $N(E_F)$. As shown in Figs. 5(d) and (i), the hole doping increases $N(E_F)$ and $\lambda$. On the other hand, electron doping decreases $N(E_F)$ but increases $\lambda$. That can be understood from Figs. 5 (e)-(g): Upon hole doping, the frequency of the mode A changes slightly, while the modes B and C are softened largely and mainly contribute to the increase of $\lambda$. On the contrary, upon electron doping, the frequencies of the modes B and C slightly change, while the mode A is significantly softened. Therefore, the electron doping increases $\lambda$ and also decreases $\omega_{log}$ remarkably (Figs. 5(h) and (i)). Therefore from Fig. 5 (j), we can conclude that the electron doping by external gated electric field might be harmful for the superconductivity, while under photoexcitation the superconductivity of borophene can be significantly enhanced. With a low doping level of 0.1 hole/cell (equal to the doped hole density of $2.16\times10^{-14}\ cm^{-2}$), the predicted $T_c$ is as high as 34.8 K. Higher doping concentration will lead to phonon instability.
\section{Conclusion}
In conclusion, based on the first principles calculations, we demonstrate that the superconductivity in borophene with vacancy-free structure could be significantly enhanced by tensile strain and hole doing. The intrinsic metallic ground state with high $N(E_F)$ and the strong Fermi surface nesting lead to the sizeable electron-phonon coupling, making the freestanding borophene superconduct with $T_c$ of 19.0 K. The tensile strain can increase $T_c$ to 27.4 K, while the hole doping can increase $T_c$ to 34.8 K, which is far more above the liquid hydrogen temperature of 20.3 K. Our predictions suggest that the superconductivity of borophene grown on a substrate with large lattice parameters or under photoexcitation could be enhanced significantly, which will largely broaden the applications of such novel material.

\begin{acknowledgments}
This work was supported by the National Key Basic Research under Contract No. 2011CBA00111, the National Nature Science Foundation of China under Contract Nos. 11404342, 11274311, U1232139, Anhui Provincial Natural Science Foundation under Contract No. 1408085MA11, and Youth Innovation Promotion Association of CAS (2012310). The calculations were partially performed at the Center for Computational Science, CASHIPS.
\end{acknowledgments}


\begin{thebibliography}{50}

\bibitem{science_666-669} K. S. Novoselov, A. K. Geim, S. V. Morozov, D. Jiang, Y. Zhang, S. V. Dubonos, I. V. Grigorieva, and A. A. Firsov, Science \textbf{306}, 666 (2004).
\bibitem{nature_197-200} K. S. Novoselov, A. K. Geim, S. V. Morozov, D. Jiang, M. I. Katsnelson, I. V. Grigorieva, S. V. Dubonos, and A. A. Firsov, Nature \textbf{438}, 197 (2005).
\bibitem{Nature_nanotechnology_699-712} Q. H. Wang, K. Kalantar-Zadeh, A. Kis, J. N. Coleman, and M. S. Strano, Nat. Nanotechnol. \textbf{7}, 699 (2012).
\bibitem{Andrea-nanolett-2010} A. Splendiani, L. Sun, Y. Zhang, T. Li, J. Kim, C.-Y. Chim, G. Galli, and F. Wang, Nano Lett. \textbf{10}, 1271 (2010).
\bibitem{PRL-Mak-2010} K. F. Mak, C. Lee, J. Hone, J. Shan, and T. F. Heinz, Phys. Rev. Lett. \textbf{105}, 136805 (2010).
\bibitem{Science_217-220} M. Corso, W. Auw\"{a}rter, M. Muntwiler, A. Tamai, T. Greber, and J. Osterwalder, Science \textbf{303}, 217 (2004).
\bibitem{Pacile-APL-2008} D. Pacil\'{e}, J. C. Meyer, C. O. Girit, and A. Zettl, Appl. Phys. Lett. \textbf{92}, 133107 (2008).
\bibitem{Song-nanolett-2010} L. Song, L. Ci, H. Lu, P. B. Sorokin, C. Jin, J. Ni, A. G. Kvashnin, D. G. Kvashnin, J. Lou, B. I. Yakobson \textit{et al.}, Nano Lett. \textbf{10}, 3209 (2010).
\bibitem{Applied_Physics_Letters_183102} B. Aufray, A. Kara, S. Vizzini, H. Oughaddou, C. L\'{e}andri, B. Ealet, and G. Le Lay, Appl. Phys. Lett. \textbf{96}, 183102 (2010).
\bibitem{Applied-Physics-Letters-223109} B. Lalmi, H. Oughaddou, H. Enriquez, A. Kara, S. Vizzini, B. Ealet, and B. Aufray, Appl. Phys. Lett. \textbf{97}, 223109 (2010).
\bibitem{Vogt-2012-PRL} P. Vogt, P. De Padova, C. Quaresima, J. Avila, E. Frantzeskakis, M. C. Asensio, A. Resta, B. Ealet, and G. Le Lay, Phys. Rev. Lett. \textbf{108}, 155501 (2012).
\bibitem{Nano_Research_853-859} W. Lu, H. Nan, J. Hong, Y. Chen, C. Zhu, Z. Liang, X. Ma, Z. Ni, C. Jin, and Z. Zhang, Nano Res. \textbf{7}, 853 (2014).
\bibitem{Nat_Nanotechnol_372-7} L. Li, Y. Yu, G. J. Ye, Q. Ge, X. Ou, H. Wu, D. Feng, X. H. Chen, and Y. Zhang, Nat. Nanotechnol. \textbf{9}, 372 (2014).
\bibitem{Nat_Nano_703-711} S. De Franceschi, L. Kouwenhoven, C. Sch\"{o}nenberger, and W. Wernsdorfer, Nat. Nano. \textbf{5}, 703 (2010).
\bibitem{Science_1045-1048} J. Delahaye, J. Hassel, R. Lindell, M. Sillanp\"{a}\"{a}, M. Paalanen, H. Sepp\"{a}, and P. Hakonen, Science \textbf{299}, 1045 (2003).
\bibitem{Applied-Physics_Letters_222506} E. J. Romans, E. J. Osley, L. Young, P. A. Warburton, and W. Li, Appl. Phys. Lett. \textbf{97}, 222506 (2010).
\bibitem{Phys_Rev_Lett_027203} O. P. Saira, M. Meschke, F. Giazotto, A. M. Savin, M. M\"{o}tt\"{o}nen, and J. P. Pekola, Phys. Rev. Lett. \textbf{99}, 027203 (2007).
\bibitem{Physical_Review_B_134530} M. Huefner, C. May, S. Ki\v{c}in, K. Ensslin, T. Ihn, M. Hilke, K. Suter, N. F. de Rooij, and U. Staufer, Phys. Rev. B \textbf{79}, 134530 (2009).
\bibitem{Nature_Physics_131-134} G. Profeta, M. Calandra, and F. Mauri, Nat. Phys. \textbf{8}, 131 (2012).
\bibitem{Guzman-2D-2014} D. M. Guzman, H. M. Alyahyaei, and R. A. Jishi, 2D Mater. \textbf{1}, 021005 (2014).
\bibitem{Physical_review_letters_037002} G. Savini, A. C. Ferrari, and F. Giustino, Phys. Rev. Lett. \textbf{105}, 037002 (2010).
\bibitem{Phys-Rev-Lett-196802} C. Si, Z. Liu, W. Duan, and F. Liu, Phys. Rev. Lett. \textbf{111}, 196802 (2013).
\bibitem{Physical_Review_B_241408} Y. Ge and A. Y. Liu, Phys. Rev. B \textbf{87}, 241408 (2013).
\bibitem{EPL_(Europhysics_Letters)_36001} W. Wan, Y. Ge, F. Yang, and Y. Yao, Europhys. Lett. \textbf{104}, 36001 (2013).
\bibitem{EPL_(Europhysics_Letters)_67004} D. F. Shao, W. J. Lu, H. Y. Lv, and Y. P. Sun, Europhys. Lett. \textbf{108}, 67004 (2014).
\bibitem{New_Journal_of_Physics_035008} Y. Ge, W. Wan, F. Yang, and Y. Yao, New J. Phys. \textbf{17}, 035008 (2015).
\bibitem{Ludbrook-PNAS-2015} B. M. Ludbrook, G. Levy, P. Nigge, M. Zonno, M. Schneider, D. J. Dvorak, C. N. Veenstra, S. Zhdanovich, D. Wong, P. Dosanjh \textit{et al.}, Proc. Natl. Acad. Sci. USA \textbf{112}, 11795 (2015).
\bibitem{Science_1513-1516} A. J. Mannix, X. F. Zhou, B. Kiraly, J. D. Wood, D. Alducin, B. D. Myers, X. Liu, B. L. Fisher, U. Santiago, J. R. Guest \textit{et al.}, Science \textbf{350}, 1513 (2015).
\bibitem{arXiv_1512.05029} B. Feng, J. Zhang, Q. Zhong, W. Li, S. Li, H. Li, P. Cheng, S. Meng, L. Chen, and K. Wu, Nat. Chem. (in press) (2016). (also see: arXiv:1512.05029 (2015))
\bibitem{Science_1468-1469} H. Sachdev, Science \textbf{350}, 1468 (2015).
\bibitem{arXiv_1602.03620} L. C. Xu, A. Du, and L. Kou, arXiv:1602.03620 (2016).
\bibitem{arxiv-1601} B. Peng, H. Zhang, H. Shao, Y. Xu, R. Zhang, and H. Zhu, arXiv:1601.00140 (2016).
\bibitem{Penev-nano-lett-2016} E. S. Penev, A. Kutana, and B. I. Yakobson, Nano Lett. (in press) (2016) .
\bibitem{arXiv_1602.02930} M. Gao, Q. Z. Li, X. W. Yan, and J. Wang, arXiv:1602.02930 (2016).
\bibitem{arxive-1512.0527} B. Feng, J. Zhang, R. Y. Liu, I. Takushi, C. Lian, L. Chen, K. Wu, H. Li, S. Meng, F. Komori \textit{et al.}, arXiv:1512.05270 (2015).
\bibitem{Arxiv-01393} S. G. Xu, Y. J. Zhao, J.-H. Liao, X.-B. Yang, and H. Xu, arXiv:1601.01393 (2016).
\bibitem{Angewandte_Chemie_13214-13218} Z. Zhang, Y. Yang, G. Gao, and B. I. Yakobson, Angewandte Chemie \textbf{127}, 13214 (2015).
\bibitem{Journal_of_Physics_Condensed_Matter_395502} P. Giannozzi, S. Baroni, N. Bonini, M. Calandra, R. Car, C. Cavazzoni, D. Ceresoli, G. L. Chiarotti, M. Cococcioni, I. Dabo \textit{et al.}, J. Phys.: Condens. Matter \textbf{21}, 395502 (2009).
\bibitem{Physical_Review_B_7892} D. Vanderbilt, Phys. Rev. B \textbf{41}, 7892 (1990).
\bibitem{Phys.-Rev.-B-6671-6687} J. P. Perdrew, J. A. Chevary, S. H. Vosko, K. A. Jacson, M. R. Pederson, D. J. Singh, and C. Fiolais, Phys. Rev. B \textbf{15}, 6671 (1992).
\bibitem{Reviews_of_Modern_Physics_515-562} S. Baroni, S. de Gironcoli, A. Dal Corso, and P. Giannozzi, Rev. Mod. Phys. \textbf{73}, 515 (2001).
\bibitem{arXiv_1602.00456} H. Wang, Q. Li, Y. Gao, F. Miao, X. F. Zhou, and X. Wan, arXiv:1602.00456 (2016).
\bibitem{Physical_Review_B_905} P. B. Allen and R. C. Dynes, Phys. Rev. B \textbf{12}, 905 (1975).
\bibitem{Physical_Review_162} J. Bardeen, L. N. Cooper, and J. R. Schrieffer, Phys. Rev. \textbf{106}, 162 (1957).
\bibitem{Physical_Review_1175} J. Bardeen, L. N. Cooper, and J. R. Schrieffer, Phys. Rev. \textbf{108}, 1175 (1957).
\bibitem{Cooper-1189} L. N. Cooper, Phys. Rev. \textbf{104}, 1189 (1956).
\bibitem{Science_1193-1196} J. T. Ye, Y. J. Zhang, R. Akashi, M. S. Bahramy, R. Arita, and Y. Iwasa, Science \textbf{338}, 1193 (2012).
\bibitem{Nat_Commun_8826} J. Biscaras, Z. Chen, A. Paradisi, and A. Shukla, Nat. Commun. \textbf{6}, 8826 (2015).
\bibitem{Science_advances_e1500168} I. Vaskivskyi, J. Gospodaric, S. Brazovskii, D. Svetin, P. Sutar, E. Goreshnik, I. A. Mihailovic, T. Mertelj, and D. Mihailovic, Science Adv. \textbf{1}, e1500168 (2015).
\bibitem{arXiv_1512.06553} D. F. Shao, R. C. Xiao, W. J. Lu, H. Y. Lv, J. Y. Li, X. B. Zhu, and Y. P. Sun, arXiv:1512.06553 (2015).
\bibitem{Science_177-180} L. Stojchevska, I. Vaskivskyi, T. Mertelj, P. Kusar, D. Svetin, S. Brazovskii, and D. Mihailovic, Science \textbf{344}, 177 (2014).
\end{thebibliography}
\end {document}